\documentclass[a4paper,11pt]{article}
\pdfoutput=1 
\usepackage{jheppub} 
\usepackage[T1]{fontenc} 
\usepackage[utf8]{inputenc}
\usepackage{xcolor}
\usepackage{slashed}
\usepackage{upgreek}
\usepackage{bigints}
\usepackage{multirow}
\usepackage{caption}
\usepackage{amsmath,amssymb,bbold,epsfig}
\usepackage{amsfonts}
\usepackage{bbm}
\usepackage{subcaption}  
\newcommand{\be}{\begin{equation}}
\newcommand{\ee}{\end{equation}}
\newcommand{\bea}{\begin{eqnarray}}
\newcommand{\eea}{\end{eqnarray}}

\definecolor{darkgreen}{HTML}{109930}

\title{\boldmath Can quantum statistics help distinguish Dirac from Majorana neutrinos?}

\author[]{Evgeny Akhmedov}
\author[]{and Andreas Trautner}

\affiliation[]{Max-Planck-Institut f\"ur Kernphysik, Saupfercheckweg 1, 69117 Heidelberg, Germany}

\emailAdd{akhmedov@mpi-hd.mpg.de}
\emailAdd{trautner@mpi-hd.mpg.de}

\abstract
{Finding out if neutrinos are Dirac or Majorana particles  
is known to be extremely difficult due to the smallness of neutrino mass 
and the fact that in the limit $m_\nu=0$ both Dirac and Majorana neutrinos 
become Weyl particles, i.e.\ are indistinguishable. There have been 
suggestions in the literature that in the case of processes with production 
of a neutrino-antineutrino pair (if neutrinos are Dirac particles) or two 
neutrinos (if they are of Majorana nature) quantum statistics may be of help.  
This is because for Majorana neutrinos quantum indistinguishability of 
identical particles requires the amplitude of the process to be 
antisymmetrized with respect to the interchange of the final-state neutrinos, 
whereas no such antisymmetrization must be done for Dirac neutrinos. 
It has been claimed that the resulting differences between the cross sections 
for Dirac and Majorana neutrinos persist even for arbitrarily small but not 
exactly vanishing neutrino mass. We demonstrate that, at least in the 
framework of the Standard Model, this is not the case. We also give a 
general proof that within the Standard Model quantum statistics does not 
help tell Dirac and Majorana neutrinos apart in the limit of negligibly small 
$m_\nu/E$.}

\begin{document} 
\maketitle
\flushbottom

\section{Introduction}
\label{sec:Introduction}
One of the most fundamental questions of neutrino physics is whether 
neutrinos are different from their antiparticles, i.e.\ are Dirac fermions, 
or are their own antiparticles, which would mean that they are of Majorana 
nature. Despite significant experimental effort, we still do not know the 
answer to this question. It is generally believed that 
the tremendous experimental difficulties in finding out the Dirac {\em vs.} 
Majorana nature of neutrinos are related to the extreme smallness of the 
neutrino mass $m_\nu$ compared to typical neutrino energies, as in the limit 
of vanishing $m_\nu$ the difference between Dirac and Majorana neutrinos 
disappears. 
 
The main distinction between neutrinos of Dirac and Majorana nature is that 
Dirac neutrinos are described by four-component spinor fields and possess a 
conserved lepton number, whereas Majorana neutrinos are described by 
two-component fields and no conserved lepton number can be ascribed to them. 
In the limit of vanishing neutrino mass the left-handed and right-handed 
components of both Dirac and Majorana neutrinos decouple and become 
two-component Weyl fields. As the 
right-handed components of Dirac neutrinos are sterile in the Standard Model, 
any distinction between the Dirac and Majorana neutrinos then disappears 
\cite{Case:1957zza,Li:1981um,Kayser:1981nw,Kayser:1982br}: in both cases, the 
states that can be produced or absorbed by weak interactions are left-handed 
neutrinos 
$\nu_L$ and their right-handed CPT conjugates $(\nu_L)^c=\nu^c_R$.  This, of 
course, need not be the case beyond the Standard Model, where the right-handed 
components of Dirac neutrinos $\nu_R$ may not be sterile (see, e.g., 
\cite{Rodejohann:2017vup}).

The difficulties with telling apart Dirac and Majorana neutrinos 
suggest looking for processes which, even if rare, are strictly forbidden 
for neutrinos of one of the two types. Currently the most promising 
candidates appear to be the neutrinoless double $\beta$-decay and related 
processes, which can only occur if neutrinos are Majorana 
particles~\cite{Schechter:1981bd} (see e.g. \cite{Dolinski:2019nrj} for a 
recent review). 
  
The statement that, at least within the framework of the Standard Model, the 
smallness of neutrino mass makes Dirac and Majorana neutrinos practically 
indistinguishable was dubbed the Practical Dirac-Majorana Confusion Theorem 
\cite{Kayser:1981nw,Kayser:1982br}. Crucial to 
it is the fact that both charged-current (CC) and neutral-current (NC) 
interactions of neutrinos in the Standard Model are purely chiral: 
$j^\mu_{\rm CC}(x)=\bar{l}(x)\gamma^\mu(1-\gamma_5)\nu_l(x)$, 
$j^\mu_{\rm NC}=\bar{\nu}_l(x)\gamma^\mu(1-\gamma_5)\nu_l(x)$.%
\footnote{Strictly speaking, the neutral current is chiral only for 
Dirac neutrinos, whereas for Majorana neutrinos it is purely axial-vector. 
However, in the limit $m_\nu/E\to 0$ this makes no difference, 
see sections~\ref{sec:neutral} and~\ref{sec:NCdetection} below.}

There have been suggestions in the literature that there may exist exceptions 
to this theorem; in particular,  it has been argued that in the case of 
processes with production of a neutrino-antineutrino pair (if neutrinos are 
Dirac particles)   or two neutrinos (if they are of Majorana nature)  
quantum statistics may help discriminate between the two neutrino types 
\cite{Ma:1989jpa,Kogo:1991ec,Hofer:1996cs,Kim:2021dyj,Kim:2023iwz}. 
The argument is based on the fact that for Majorana neutrinos 
quantum indistinguishability of identical particles requires the amplitude of 
the process to be antisymmetrized with respect to the interchange of the 
final-state neutrinos, whereas no such antisymmetrization must be done if 
neutrinos are Dirac particles, as the produced $\nu$ and $\bar{\nu}$ are 
distinct. It has been claimed  that the resulting differences between the 
cross sections for Dirac and Majorana neutrinos survive even for arbitrarily 
small but not exactly vanishing neutrino mass, though they disappear 
when $m_\nu=0$.  

Such a lack of smooth behaviour of the cross sections in the limit 
$m_\nu\to 0$ is very counterintuitive and unsettling, as one naturally 
expects physical observables to be continuous 
functions of the masses of the involved fermions. 
In the present paper we examine this issue in detail. We demonstrate that the 
claims of the absence of smooth behaviour in the limit $m_\nu\to 0$ 
do not hold, as far as observable quantities are concerned. 
We also prove that this result is not limited to the processes considered 
in \cite{Ma:1989jpa,Kogo:1991ec, Hofer:1996cs,
Kim:2021dyj,Kim:2023iwz}, but holds true for all Standard Model processes.%
\footnote{A qualification is in order. 
To be precise, neutrinos are massless in the 
minimal Standard Model; what we consider in this paper is actually 
``the Standard Model plus neutrino mass'', i.e.\ we assume that 
neutrino mass is generated by an unspecified new physics (possibly at a high 
energy scale) that does not affect low-energy processes, except through the 
neutrino mass itself. 
We also note that leptonic mixing 
(and more generally the existence of more than one neutrino flavor) is  
not directly relevant to the issues we consider, and so we disregard it 
in our discussion. }
Our conclusion is thus that quantum statistics does not lead to any exceptions 
to the Practical Dirac-Majorana Confusion Theorem. 

The paper is organized as follows. 
In section~\ref{sec:general} we first review the well known examples 
illustrating how the Practical Dirac-Majorana Confusion Theorem works in the 
cases of decay/inverse decay and neutrino scattering processes. These examples 
demonstrate that for Majorana neutrinos the role of the (nearly) conserved 
lepton number is played by chirality, which is approximately conserved for 
relativistic neutrinos. 
We then give very general arguments for why this breaks 
the indistinguishability of Majorana 
neutrinos in processes with their pair-production.   
In sections~\ref{sec:confusion} and \ref{sec:solution} 
we discuss the existing claims that for 
processes with two neutrinos in the final state quantum statistics leads to 
differences between the cross sections for Dirac and Majorana neutrinos 
that do not disappear in the limit $m_\nu\to 0$ and demonstrate 
that these claims are erroneous. In section~\ref{sec:generalProof} we 
present a general proof that quantum statistics does not 
violate the Practical Dirac-Majorana Confusion Theorem. 
The reader content with general arguments and not interested in details of 
our analysis may skip sections \ref{sec:confusion}, \ref{sec:solution} and 
\ref{sec:generalProof} and go directly to section~\ref{sec:disc}, where we 
summarize our results and conclude.

\section{General arguments}
\label{sec:general}
\subsection{\label{sec:1nu}Single neutrino production, absorption and 
scattering in CC processes}
One well-known example of how the Practical Dirac-Majorana Confusion Theorem 
works is given by $\beta$-decay and inverse $\beta$-decay processes, which 
are induced by the CC weak interactions (see, e.g., 
ref.~\cite{Bilenky:2006we}). 

It is known that electron neutrinos $\nu_e$ produced in $\beta^+$-decays, 
(e.g.\ solar neutrinos) are different from those produced in $\beta^-$-decays, 
which are usually called $\bar{\nu}_e$. The former can be detected through the 
reaction $\nu_e+{\rm ^{37}Cl}\to {\rm^{37}Ar}+e^-$, 
whereas the detection of these neutrinos through the inverse $\beta$-decay 
on protons has never been observed. On the other hand, 
the latter (produced e.g.\ in $\beta$-decays of heavy nuclei in 
nuclear reactors) 
can be (and usually are) detected through the inverse $\beta$-decay 
on protons, whereas the attempts to detect them through the Cl-Ar reaction 
mentioned above have failed. 
These facts can be easily explained if neutrinos are Dirac particles and 
possess conserved lepton number. In this case $\nu_e$ and $\bar{\nu}_e$ are 
electron neutrinos and antineutrinos, respectively, and the selection rules 
discussed above are just a consequence of lepton number conservation. 

Does this mean that we already have an experimental proof that neutrinos are 
different from their antiparticles, i.e.\ they are Dirac fermions? 
The answer is no, of course. The point is that the chiral structure of  
the weak currents means that leptons participate in CC 
weak interactions only by their left-handed chirality components, 
and antileptons by their right-handed chirality components.
Chirality is not a good quantum number for fermions of nonzero mass, but for 
free relativistic particles chirality nearly coincides with helicity, which 
{\sl is} conserved; the difference between the two is of the order of 
$m_\nu/E$. For $u$-type and $v$-type spinors describing, respectively, the 
positive-energy and negative-energy solutions of the Dirac 
equation, we have 
\begin{align}
u_L(p)\simeq u_-(p)+{\cal O}\Big(\frac{m_\nu}{2E}\Big)
\,,\qquad 
v_R(p)\simeq v_+(p)+{\cal O}\Big(\frac{m_\nu}{2E}\Big)
\,,
\nonumber \\
u_R(p)\simeq u_+(p)+{\cal O}\Big(\frac{m_\nu}{2E}\Big)
\,,\qquad 
v_L(p)\simeq v_-(p)+{\cal O}\Big(\frac{m_\nu}{2E}\Big)
\,,
\label{eq:uv}
\end{align}
where 
$u_{L,R}=P_{L,R}u$, $v_{L,R}=P_{R,L} v$ with $P_{L,R}=\frac{1}{2}(1\mp 
\gamma_5)$, and the subscripts $\pm$ stand for positive and negative 
helicities. 
Eq.~(\ref{eq:uv}) means that relativistic neutrinos produced together with 
positively charged leptons (or absorbed in reactions with production of 
negatively charged leptons) 
are predominantly negative helicity states, whereas neutrinos 
produced together with negatively charged leptons (or absorbed in reactions 
with production of positively charged leptons) 
are predominantly states of positive helicity. 
Thus, the chirality selection rules of CC weak interactions 
play essentially the same role for relativistic Majorana neutrinos 
as lepton number conservation plays for Dirac neutrinos. 
For example, what we call $\bar{\nu}_e$ is an electron 
antineutrino in the Dirac case and an electron neutrino 
of nearly positive helicity 
if neutrinos are Majorana particles.
The difference is that chirality is 
only approximately conserved for relativistic neutrinos with $m_\nu\ne 0$, 
whereas for Dirac neutrinos lepton number is conserved exactly. 
Therefore, in the Majorana case the processes like detection of solar neutrinos 
through inverse $\beta$-decay on protons (and other ``wrong-helicity'' 
processes) are not strictly forbidden, but are strongly suppressed; 
the suppression factors are ${\cal O}(m_\nu/(2E))^2\lesssim 10^{-14}$, 
which explains why such processes have never been observed. 

Obviously, the same arguments apply also to other decay and inverse decay   
CC processes, including those with participation of $\nu_\mu$ or $\nu_\tau$. 
They are also valid for CC contributions to $\nu_\ell \ell$ scattering, 
where the same chirality selection rules are at work.

\subsection{\label{sec:neutral}Neutral current induced scattering processes}

What about neutral-current neutrino interactions? 
There is a peculiarity in this case.  
For Majorana particles the vector neutral current vanishes identically: 
$\bar{\psi}(x)\gamma^\mu \psi(x)\equiv 0$. This can be easily proven by 
making use of the self-conjugacy property of Majorana fields $(\psi)^c\equiv 
C\bar{\psi}^T=\psi$, where $C$ is the charge conjugation matrix. As a result, 
for Majorana neutrinos the neutral current is purely axial-vector: 
\be
\bar{\nu}(x)\gamma^\mu(g_V-g_A\gamma_5)\nu(x)=-g_A\bar{\nu}(x)
\gamma^\mu\gamma_5\nu(x)\,,
\label{eq:axial}
\ee
where we have introduced the vector and axial-vector coupling constants $g_V$ 
and $g_A$ for future convenience. 
The measurement of the 
NC $\nu_\mu e \to \nu_\mu e$ and  
$\bar{\nu}_\mu e \to \bar{\nu}_\mu e$ scattering cross sections by 
the CHARM-II collaboration \cite{CHARM-II:1993xmq,CHARM-II:1994dzw} was 
interpreted in \cite{Plaga:1996it} as an experimental evidence for  
nonzero vector neutrino NC coupling $g_V$. The author then concluded that 
neutrinos cannot be Majorana particles. It has been subsequently demonstrated 
in~\cite{Kayser:1997hj,Hannestad:1997mi,Hansen:1997sk,Zralek:1997sa,Czakon:1999ed} 
that this interpretation was incorrect. The amplitude of the process depends 
on the NC matrix element $\langle \nu(p')|j^\mu_{\rm NC}(0)|\nu(p)\rangle$, 
for which in the Majorana case one finds 
\begin{align}  
&\bar{u}(p')\gamma^\mu(g_V-g_A\gamma_5) u(p)-\bar{v}(p)\gamma^\mu
(g_V-g_A\gamma_5) v(p')=\nonumber \\
&\bar{u}(p')\gamma^\mu(g_V-g_A\gamma_5) u(p)-\bar{u}(p')\gamma^\mu
(g_V+g_A\gamma_5) u(p)=-2g_A\bar{u}(p')\gamma^\mu\gamma_5u(p)\,.
\label{eq:MajNeutral1}
\end{align}
Here $p$ and $p'$ are the 4-momenta of the initial- and final-state neutrinos, 
and in going from the first to the second line 
use has been made of the identities%
\footnote{Transformations similar to that in 
eq.~(\ref{eq:MajNeutral1}) are also made in a number of equations appearing 
below.}
\be
u_s(p)=C\bar{v}_s^T(p)\,,\qquad v_s(p)=C\bar{u}_s^T(p)\,.
\label{eq:ident}
\ee
Thus, for Majorana neutrinos the NC matrix element is purely 
axial-vector, as expected. 

In the Dirac case, the matrix element of the neutrino neutral current is 
\be
\bar{u}(p')\gamma^\mu(g_V-g_A\gamma_5) u(p)=-(g_V+g_A)\bar{u}(p')\gamma^\mu
\gamma_5 u(p)+{\cal O}\Big(\frac{m_\nu}{2E}\Big)\,, 
\label{eq:DirNeutral}
\ee
where it was taken into account that for relativistic Dirac 
neutrinos $\gamma_5 u\simeq -u$. Because in the 
Standard Model $g_V=g_A$, 
the right-hand sides of 
eqs.~(\ref{eq:MajNeutral1}) and (\ref{eq:DirNeutral})
coincide up to terms of the order of $m_\nu/(2E)$, 
which proves the Practical Dirac-Majorana Confusion Theorem for NC 
$\nu e$-scattering \cite{Kayser:1981nw,Kayser:1997hj,Hannestad:1997mi,
Hansen:1997sk,Zralek:1997sa,Czakon:1999ed}. It is easy to see 
that this result also applies to other 
NC induced neutrino scattering processes. 

Important to the above argument was the fact that the incident $\nu_\mu$ and 
what we call $\bar{\nu}_\mu$ in the NC neutrino-electron scattering experiments 
were born in CC processes \mbox{($\pi^\pm$-decays)} as left-chiral and 
right-chiral states, respectively~\cite{Kayser:1981nw}.%
\footnote{Actually, in {\sl all} neutrino detection experiments carried out so 
far (except possibly in supernova neutrino experiments) the incident neutrinos 
were produced in CC reactions.} 
Because chirality is nearly conserved for free relativistic fermions, and  
the axial-vector interaction does not flip it, the scattered neutrinos in the 
final state have essentially the same chirality as the incident ones for 
Majorana neutrinos, just as is the case for the Dirac ones. 

Thus, similarly to the case of CC reactions discussed in section~\ref{sec:1nu}, 
the crucial role in practical indistinguishability of Dirac and Majorana 
neutrinos in the NC scattering experiments with incident neutrinos produced in 
CC processes is played 
by the chiral nature of neutrino interactions. 
In section~\ref{sec:NCdetection} we shall discuss NC-induced 
scattering in the case when the incident neutrinos are 
produced in NC processes (which means that in the Majorana case their 
chirality is in general undefined), and we shall show that 
the confusion theorem remains valid in that case as well. 

\subsection{\label{sec:genArguments}Processes with two or more neutrinos 
in the final state}

Above, we discussed processes with no more than one neutrino in the final 
state. Processes with pair-production of $\nu\bar{\nu}$ or $\nu\nu$ (for 
Dirac or Majorana neutrinos, respectively) may require special consideration 
because of quantum indistinguishability of the two neutrinos in the latter case. 
We shall argue now that this will not lead to any exceptions to the Practical 
Dirac-Majorana Confusion Theorem. An explicit proof of this statement will 
be given in sections~\ref{sec:solution} and~\ref{sec:generalProof}. 

It is true that Majorana neutrinos born in pair-production processes (such as 
e.g.\ $\ell^+\ell^-\to \nu\nu$) are identical, no matter how small their mass is; 
therefore, strictly speaking, the amplitude of the process must always be 
antisymmetrized with respect to their interchange. However, one can expect 
that with decreasing $m_\nu$ the {\sl observable} effects of this 
antisymmetrization will decrease, and they will become unmeasurable for 
arbitrarily small neutrino mass. Indeed, with decreasing $m_\nu$ the 
left-handed and right-handed components of the Majorana neutrino field, 
$\nu_L$ and $(\nu_L)^c\equiv\nu^c_R$, become less strongly coupled to each other,  
and the transitions between them get suppressed. 
In the limit $m_\nu/E\to 0$ they decouple and behave effectively as distinct 
particles, and the amplitude of their pair production need not be 
antisymmetrized.Technically, this should manifest itself as the 
suppression of the observable effects of the antisymmetrization by positive 
powers of $m_\nu/E$. This is actually related to the fact that for 
relativistic Majorana neutrinos chirality plays the role of an approximately 
conserved lepton number, as was pointed out above. 
All effects of chirality nonconservation should be suppressed for small 
$m_\nu/E$; manifestations of the identical nature of Majorana neutrinos in 
$\nu \nu$ pair production is just one of these effects. Clearly, the same 
argument also applies to processes with production of more than two 
Majorana neutrinos. 

\section{\label{sec:confusion}Confusion over the Dirac-Majorana confusion} 

As was mentioned above, there had been claims in the literature that in 
processes with pair-production of neutrinos the differences between the cross 
sections for Dirac and Majorana neutrinos survive in the limit $m_\nu\to 0$, 
contrary to the general arguments given in the previous section. 
In refs.~\cite{Ma:1989jpa,Kogo:1991ec,Hofer:1996cs}, this was found to occur 
for the NC processes $e^+e^-\to Z^*\to \nu\bar{\nu}(\nu\nu)$. It was pointed 
out, however, that the Dirac/Majorana differences do smoothly disappear 
in the limit $m_\nu\to 0$ if the produced neutrinos are not detected 
(or, more generally, if their spins are not measured). 
 
In refs.~\cite{Kim:2021dyj,Kim:2023iwz} the four-body decays 
$B^0\to\mu^+\mu^-\nu_\mu\bar{\nu}_\mu$ and $B^0\to\mu^+\mu^-\nu_\mu\nu_\mu$ 
mediated by second order CC interactions were considered. 
The authors concluded that for the special case of back-to-back kinematics for 
the produced muons in the rest frame of the parent $B^0$-meson 
(which by momentum conservation also implies back-to-back kinematics for the 
produced neutrinos) the differences between the differential decay rates in 
the Dirac and Majorana cases do not disappear for arbitrarily small but 
nonzero neutrino mass. Surprisingly, this was found to happen even though the 
authors studied the case when the final-state neutrinos are not detected. 
This is in contrast with the mentioned above results 
of~\cite{Ma:1989jpa,Kogo:1991ec,Hofer:1996cs} 
for neutrino pair production in 
NC processes. The authors of~\cite{Kim:2021dyj,Kim:2023iwz} argued that it was precisely 
their use of a CC process that 
led to this result, because for NC-induced processes the summation 
over the spins of unobserved neutrinos would kill  
the Dirac/Majorana difference in the limit $m_\nu/E\to 0$ \cite{Kim:2023iwz}. 

We shall now comment on the 
above results, starting with those of 
refs.~\cite{Kim:2021dyj,Kim:2023iwz}. 

\subsection{\label{sec:B0decay}Charged-current decays 
\texorpdfstring{$B^0\to\mu^+\mu^-\nu_\mu\bar{\nu}_\mu$}{B0->mu mu nu bar(nu)}
and
\texorpdfstring{$B^0\to\mu^+\mu^-\nu_\mu\nu_\mu$}{B0->mu mu nu nu}}
Unfortunately, the papers \cite{Kim:2021dyj,Kim:2023iwz}, where these processes 
were considered, are incorrect.  
This can be seen from eqs.~(48a) and (48b) of~\cite{Kim:2021dyj}, 
from which the authors deduce their main conclusions.   
These expressions present their results for triply differential 
$B^0$-meson decay rates 
for the back-to-back kinematics 
(and for negligibly small but nonzero $m_\nu$) in the cases 
of Dirac and Majorana neutrinos, respectively. Their final results are then 
obtained upon the integration over the unobservable angle $\theta$ between the 
directions of the produced neutrinos and muons. However, the denominators of 
the left-hand sides of eqs.~(48a) and (48b) erroneously contain 
$\mathrm{d}\!\sin\theta$ instead of $\mathrm{d}\!\cos\theta$. 
The authors arrived at this 
result by first defining all the angles characterizing the process in general 
kinematics and then expressing the angle $\theta$, corresponding to the 
back-to-back kinematics, through one of these angles, $\theta_m$. 
However, the transition to this special case means that the angles 
$\theta_m$ and $\theta_n$, defined for general kinematics, become unphysical. 
This can be seen from fig.~4 of~\cite{Kim:2021dyj}: back-to-back kinematics 
corresponds to vanishing vectors $\vec{q}_m$ and $\vec{q}_n$, which means that 
their directions are undefined. To overcome this difficulty, the authors define 
the physical angle $\theta$ through a limiting procedure, which, however, is 
both ambiguous and unnecessary. 

One can consider the problem in the back-to-back kinematics from the outset, 
without any limiting procedures. In this case the process is fully 
characterized by just two physical variables -- the muon energy $E_\mu$ and  
the angle $\theta$ between the muon and neutrino directions. 
Then, since the neutrinos are not
observed, one must integrate 
over the solid angle of the neutrino direction, which involves the integration 
over $\mathrm{d}\!\cos\theta$, not over $\mathrm{d}\!\sin\theta$. 
Once the correct angular integration 
of eqs.~(48a) and (48b) of~\cite{Kim:2021dyj} has been carried out, their 
right-hand sides yield identical results. This  
means that for negligibly small $m_\nu$ the differential decay rates 
for Dirac and Majorana neutrinos coincide, in full agreement with the practical 
confusion theorem. The limit $m_\nu\to 0$ is smooth, as it should be. 

\subsection{\label{sec:Znunu}Neutrino pair production in neutral-current 
processes}

We now turn to refs.~\cite{Ma:1989jpa,Kogo:1991ec,Hofer:1996cs}, 
in which the processes $e^+e^-\to Z^*\to \nu\bar{\nu}(\nu\nu)$ were 
considered. These papers are technically correct,%
\footnote{The expression for the differential cross section in the Dirac case 
$\mathrm{d}\sigma^\mathrm{D}/\mathrm{d}\Omega$ in eq.~(3) of \cite{Ma:1989jpa} contains 
some inaccuracies, which have been corrected in~\cite{Kogo:1991ec,Hofer:1996cs}
~(see eq.~(4A) of~\cite{Kogo:1991ec} or eq.~(A5) of~\cite{Hofer:1996cs}). 
Note that the limit $m_\nu\to 0$ of this cross section given in eq.~(4) 
of~\cite{Ma:1989jpa} is correct.} 
but they contain some questionable and confusing statements, see below.
The main focus in these papers was on the production of hypothetical heavy 
neutrinos and their subsequent decay; however, a large part of their results 
also applies to the production of the usual light neutrinos of the Standard 
Model, and the authors did briefly discuss the limit of very small neutrino 
masses. 

For pair-production of Dirac neutrinos, it was found 
in~\cite{Ma:1989jpa,Kogo:1991ec,Hofer:1996cs} that the differential cross 
section of the process contained, along with the usual terms that were either 
neutrino spin independent or proportional to the longitudinal components of the 
spins of the produced $\nu$ and $\bar{\nu}$, also terms that were proportional 
to their transverse spin components. However, the latter entered with the 
factors $m_\nu/E$ or $(m_\nu/E)^2$, and in the limit $m_\nu\to 0$ the standard 
expression for massless neutrinos was recovered: 
\be
\left.\frac{\mathrm{d}\sigma^\mathrm{D}}{\mathrm{d}\Omega}\right|_{m_\nu\to 0}
=\frac{\sigma_0}{2}\big[f_1(1+\cos^2\theta)+2f_2\cos\theta
\big](1-n_z)(1-n_z')\,. 
\label{eq:Dir1}
\ee
Here the coordinates are chosen such that the neutrino and antineutrino momenta 
point in the positive and negative directions of the $z$-axis, respectively 
(in the c.m.\ frame), $n_z$ and $n_z'$ are the $z$-components of the neutrino 
and antineutrino unit spin vectors 
defined in their respective rest frames, and $\theta$ is the 
angle which the momentum of the incident $e^-$, chosen to lie in the 
$xz$-plane, makes with the $z$-axis. The parameters $\sigma_0$, $f_1$ and $f_2$ 
are defined in \cite{Ma:1989jpa,Kogo:1991ec,Hofer:1996cs} and are unimportant 
to us here.\ The expression in eq.~(\ref{eq:Dir1}) is in full agreement with 
the fact that in the massless limit neutrinos become helicity eigenstates, 
i.e.\ their spins have only longitudinal components. Moreover, as expected, 
the cross section is only different from zero when 
the produced neutrino has negative helicity and the antineutrino has positive 
helicity, so that they both have spin projection $n_z=n'_z=-1$ on the 
direction of the neutrino momentum (which is the positive $z$-axis in our 
convention).  
 
At the same time, 
in the Majorana neutrino case it was found that, for arbitrary $m_\nu$,  
\be
\!\frac{\mathrm{d}\sigma^\mathrm{M}}{\mathrm{d}\Omega}=\frac{\sigma_0}{2}\beta^3\!\left\{f_1[(1+n_zn_z')
(1+\cos^2\theta)-(n_xn_x'-n_yn_y')\sin^2\theta]-2f_2(n_z+n_z')\cos\theta
\right\}\,, 
\label{eq:Maj1}
\ee
where $\beta$ is the neutrino velocity 
(see eqs.~(2), (4B) or (A6) of refs.~\cite{Ma:1989jpa}, 
\cite{Kogo:1991ec} or \cite{Hofer:1996cs}, respectively).   
The coordinate convention here is such that the produced neutrinos fly away 
in opposite directions along the $z$-axis.%
\footnote{Note that because of the indistinguishability of Majorana neutrinos 
one cannot choose the $z$-axis using the same convention as in the Dirac 
neutrino case. However, for the same reason it does not matter which of the 
produced Majorana neutrinos goes in the positive and which in the negative 
$z$-direction. It is easy to see that eq.~(\ref{eq:Maj1}) is invariant with 
respect to the interchange of the two neutrinos (which amounts to 
the coordinate transformation $z\to -z$), as it should.} 
The expression in eq.~(\ref{eq:Maj1}) contains the term 
$\propto n_xn_x'-n_yn_y'$ that depends on the transverse spin components of 
the two produced neutrinos; its origin can be traced back to the 
antisymmetrization of the amplitude of the process required by quantum 
statistics in the Majorana neutrino case. This term is not suppressed by 
positive powers of $m_\nu/E$ and therefore survives in the limit $m_\nu\to 0$. 
Since massless fermions can only have longitudinal spin components, 
the authors argue that for $m_\nu=0$ this term is unphysical and must be 
dropped. Eqs.~(\ref{eq:Dir1}) and (\ref{eq:Maj1}) then yield 
identical results.%
\footnote{To see this, 
one should remember the mentioned above 
difference in the definitions of the $z$-axis in the Dirac and 
Majorana cases, which implies that one has to flip the $z$-axis 
$(z\rightarrow -z)$ for $n_z=n'_z=1$ in the Majorana case in order to 
compare it to the Dirac case. 
See the discussion below eq.~(5) of ref.~\cite{Ma:1989jpa}.} 

The presence of an unsuppressed term depending on the transverse neutrino 
spin components even in the case of arbitrarily small 
neutrino mass is quite disturbing, and the  
prescription of dropping this   
(or actually any) 
term by hand in the case $m_\nu=0$ is unsatisfactory in our opinion. 
One naturally expects the vanishing of the contributions 
of the transverse neutrino spin components 
to be an automatic outcome of the $m_\nu\to 0$ limit, 
just as it happens in the Dirac neutrino case. 

\section{\label{sec:solution}Neutral-current neutrino pair production: 
Solution to the problem
}

In fact, a hint of how this problem can be resolved is 
already contained in refs.~\cite{Ma:1989jpa,Kogo:1991ec,Hofer:1996cs}.  
The ``anomalous'' term 
$\propto n_xn_x'-n_yn_y'$ in eq.~(\ref{eq:Maj1}) 
may only manifest itself when 
both of the final-state neutrinos are observed and their spins are measured.
This is because the summation over even one of the spins  
makes this term vanish.
Therefore, to test the effects of 
such a term one must include the neutrino observation process 
into the consideration. The authors 
of~\cite{Ma:1989jpa,Kogo:1991ec,Hofer:1996cs} 
do consider the detection of the produced neutrinos -- through their decay; 
it is known that decay processes can 
discriminate between Dirac and Majorana neutrinos (see, e.g., 
ref.~\cite{Balantekin:2018ukw} and references 
therein). However, observation of neutrinos through their decay cannot resolve 
the problem of non-smooth 
behavior of the cross sections in the $m_\nu\to 0$ 
limit, as for vanishingly small 
$m_\nu$ neutrinos become essentially stable. 
Therefore, one should consider other processes of neutrino detection, 
those that do not require finite neutrino mass.  
Within the Standard Model, 
these are only the processes related to neutrino gauge interactions, 
either 
CC or NC.

\subsection{\label{sec:CCdetection}Neutrino detection through CC processes}

\begin{figure}[t]
\begin{subfigure}[b]{0.2\textwidth}
\centering
\includegraphics[width=1.0\textwidth]{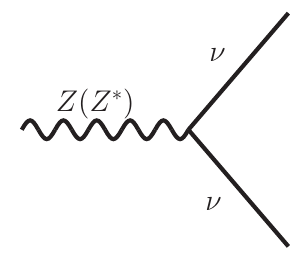}
\vspace*{8mm}
\begin{minipage}[t]{.1cm}
\vfill
\end{minipage}
\caption{\label{fig:1a}}
\end{subfigure}\hfill%
\begin{subfigure}[b]{0.33\textwidth}
\strut\vspace*{-\baselineskip}\newline
\centering
\includegraphics[width=1.0\textwidth]{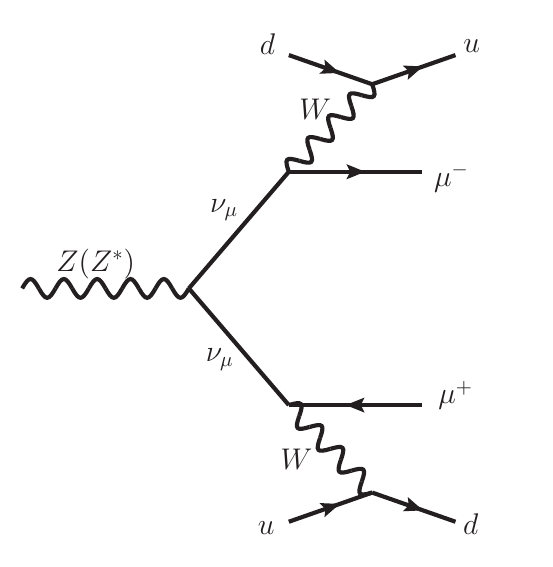}
\caption{\label{fig:1b}}
\end{subfigure}\hfill%
\begin{subfigure}[b]{0.33\textwidth}
\vspace*{-4cm}
\strut\vspace*{-\baselineskip}\newline
\centering
\includegraphics[width=1.0\textwidth]{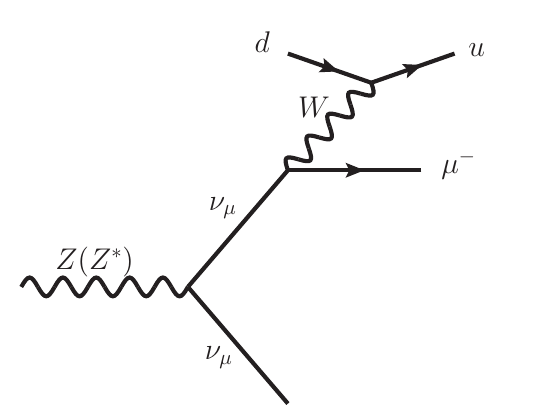}
\vspace{0.8cm}
\caption{\label{fig:1c}}
\end{subfigure}
\caption{\label{fig:1}
Feynman diagrams for some of the discussed processes. (a): production of 
a pair of Majorana neutrinos in decay of on-shell or off-shell $Z^0$-boson; 
(b) same as in (a) but with detection of both produced neutrinos (assumed to 
be of muon flavor) through CC reactions with 
production of $\mu^-$ and $\mu^+$; (c) same as in (b) but with detection of 
only one of the produced neutrinos.
}
\end{figure}

We start with neutrino detection through a 
CC process, for which we choose neutrino-nucleon (or neutrino-nucleus) 
reaction with the production of 
a charged lepton. 
To be specific, 
we consider the case of $\nu_\mu$ detection processes with $\mu^{\pm}$ in the 
final states (see fig.~\ref{fig:1b}). We only discuss  the production of 
opposite-sign muons, the reason being that the same-sign charged lepton 
production is a $\Delta L=2$ process, which means that its amplitude is 
explicitly proportional to the Majorana neutrino mass and vanishes in the limit 
$m_\nu\to 0$. 
We ignore here the experimental difficulties of simultaneous detection of two 
neutrinos from the decay of the same on-shell or off-shell 
$Z$-boson;
because we are interested in fundamental questions of 
distinguishability of Dirac and Majorana neutrinos, 
it suffices for the purposes of our study that such a 
detection is possible in principle. We shall also consider the 
situation in which only one of the produced neutrinos is detected, whereas 
the other escapes unobserved (see fig.~\ref{fig:1c}). 

As can be seen from the diagrams of figs.~\ref{fig:1b} and \ref{fig:1c}, 
the difficulty with the ``anomalous'' term in the cross section in 
eq.~(\ref{eq:Maj1}), which was an outcome of the antisymmetrization of the 
amplitude in the Majorana case, is immediately resolved by neutrino detection: 
for the processes shown in these figures 
there are either no neutrinos or only one neutrino 
in the final state, so 
there is nothing to antisymmetrize. 

Thus, there are two possibilities: (1) The neutrinos produced in 
the $e^+e^-\to Z^*\to \nu\nu$ process are not detected. The ``anomalous'' 
term in eq.~(\ref{eq:Maj1}) is then averaged away because of the 
summation of the spins of the unobserved neutrinos. (2) Either one 
or both of the final-state neutrinos are detected. The ``anomalous'' term, 
which originates from the antisymmetrization procedure,  
then does not appear at all because there are no pairs of identical neutrinos 
in the final state. Thus, we come to the conclusion that 
terms in the cross sections of 
the $e^+e^-\to Z^*\to \nu\bar{\nu}(\nu\nu)$ process 
that are different in the case of Dirac 
and Majorana neutrinos and are not 
suppressed by powers of $m_\nu/E$ never appear, as far as observable 
quantities are concerned. It is not necessary to drop any terms from the 
cross sections by hand in the case of exactly vanishing $m_\nu$. 
There are, of course, still terms that are different in the cases of Dirac and 
Majorana neutrinos, but they are all suppressed by factors $m_\nu/E$ and 
$(m_\nu/E)^2$ and thus do not violate the Practical Dirac-Majorana Confusion 
Theorem.

The above argument is very simple and is valid independently 
of the distance between the neutrino production and detection points. It is, 
however, instructive to consider separately 
the situation when the production 
and detection processes are separated by macroscopically large distances 
and therefore can be considered as independent.%
\footnote{See e.g. the discussion in section 4 of \cite{Akhmedov:2010ms}.}
This will allow us to find out how the ``anomalous'' term in 
eq.~(\ref{eq:Maj1}) actually becomes inoperative when at least one of the 
produced neutrinos is detected. 

Let us first note that the pure axial-vector nature of the neutrino 
NC in the Majorana case is actually a 
result of the coherent superposition of the left-chiral and 
right-chiral contributions with equal weights. Indeed, the amplitude of 
the $e^+e^-\to Z^*\to \nu\nu$ process 
depends on the matrix element 
$\langle \nu_{s_1}(p_1)\nu_{s_2}(p_2)|j^\mu_{\rm NC}(0)|0\rangle$, 
which is given by%
\footnote{
Note the similarity with eq.~(\ref{eq:MajNeutral1}).}
\begin{align}\label{eq:MajNeutral2}
&
\frac{1}{\sqrt{2}} \left[
\bar{u}_{s_1}(p_1)\gamma^\mu(1-\gamma_5)v_{s_2}(p_2)-
\bar{u}_{s_2}(p_2)\gamma^\mu(1-\gamma_5)v_{s_1}(p_1)\right]=
\\ \nonumber
&\frac{1}{\sqrt{2}}\left[\bar{u}_{s_1}(p_1)\gamma^\mu(1-\gamma_5)v_{s_2}(p_2)
-\bar{u}_{s_1}(p_1)\gamma^\mu(1+\gamma_5)v_{s_2}(p_2)\right]
=-\sqrt{2}\bar{u}_{s_1}(p_1)\gamma^\mu\gamma_5 v_{s_2}(p_2)\,. 
\end{align}
Here 
$p_1,s_1$ and $p_2,s_2$ are the four-momenta and the spin indices of the two 
produced neutrinos, and the factor  $1/\sqrt{2}$
comes from the 
identical nature of the two Majorana neutrinos in the final state. 

Let us now return to the process shown in fig.~\ref{fig:1b}. From the 
chirality selection rules of CC interactions discussed in section~\ref{sec:1nu} 
it follows that the neutrino producing $\mu^-$ must be of predominantly 
negative-helicity, while the neutrino producing $\mu^+$ must have 
predominantly positive helicity; in the limit $m_\nu\to 0$ they become pure 
helicity eigenstates. Thus, in this limit the considered neutrino detection 
process breaks the indistinguishability of the two Majorana neutrinos 
produced in the $e^+e^-\to Z^*\to \nu\nu$ reaction. This, in particular, means 
that for negligibly small neutrino mass there will be 
no contribution to the squared matrix element of the process from the  
interference of the two terms in the first line of eq.~(\ref{eq:MajNeutral2}) 
(or, equivalently, of the two terms on the left-hand side of the second line). 
It should be noted that the 
$(n_xn_x'-n_yn_y')$ 
term in eq.~(\ref{eq:Maj1}) comes precisely from this interference.  
Therefore, 
for $m_\nu\to 0$ 
there will be no contribution of the transverse neutrino spin components to 
the observables, 
and one does not need to drop this term by hand. 
For $m_\nu\neq 0$, the contribution of the interference term to the cross 
section of the process will be different from zero to the same extent to which 
chirality deviates from helicity, i.e.\ it will be suppressed by positive 
powers of $m_\nu/E$ (see eqs.~(\ref{eq:nueMaj3})-(\ref{eq:Tsum}) below 
and the discussion around them). 

Keeping only the squared moduli of the two terms in the second line of 
eq.~(\ref{eq:MajNeutral2}) 
would reproduce the squared matrix element of the 
overall process of fig.~\ref{fig:1b} for Dirac neutrinos, with $\mu^-$ detection 
in a given detector 
and $\mu^+$ production in the other one. 
In the Majorana case, there is an additional factor of 2 due to the fact 
that each of the two detectors can observe $\mu^-$, with the other 
observing $\mu^+$;%
\footnote{There is no such freedom in the Dirac case due to the 
specific choice of the positive direction of the $z$ axis that led to 
eq.~(\ref{eq:Dir1}).} 
with the factor $1/\sqrt{2}$ in eq.~(\ref{eq:MajNeutral2}) taken into 
account, this means that the cross sections for Dirac and Majorana neutrinos  
coincide in the massless neutrino limit. 

Thus, the detection process breaks the indistinguishability of the produced 
Majorana neutrinos in this case, which means that no antisymmetrization 
needs to be done (or, more precisely, that the antisymmetrization 
does not lead to any observable effects) in the case of negligible neutrino 
mass. 
Moreover, 
the indistinguishability of the two Majorana neutrinos produced in 
$e^+e^-\to Z^*\to \nu\nu$ reaction is broken  
even if only one of them is detected (see fig.~\ref{fig:1c}): this would 
single out the predominant helicity of the detected neutrino, and the other 
one would automatically have the opposite predominant helicity due to 
their entanglement. 
This can be immediately seen from eq.~(\ref{eq:MajNeutral2}): since 
$P_L=\frac{1}{2}(1-\gamma_5)$ projects out left-handed components of 
$u$-type spinors and right-handed components of $v$-type spinors, the first 
term in the first line of~(\ref{eq:MajNeutral2}) corresponds to neutrino of 
momentum $p_1$ being left-chiral and that of momentum $p_2$ right-chiral, 
while the second term corresponds to the opposite situation. Thus the 
detection process, by selecting the chirality of the observed neutrino, 
determines also the chirality of the other one. 

\subsection{\label{sec:NCdetection}Neutrino detection through NC 
\texorpdfstring{$\nu_\mu e$}{neutrino electron} scattering}

\begin{figure}[t]
\centering
\includegraphics[width=0.33\textwidth]{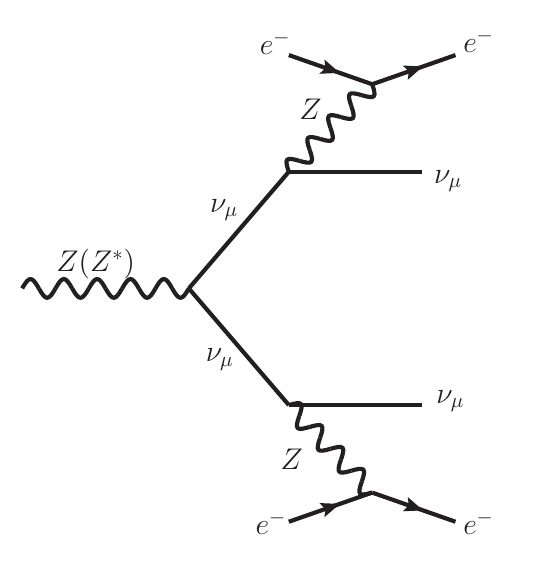}
\caption{\label{fig:2}
Same as in fig.~\ref{fig:1b} but for neutrino detection through NC $\nu_\mu e$ 
scattering.}
\end{figure}

We now turn to neutrino detection via NC process, for which we consider 
$\nu_\mu e$ scattering (fig.~\ref{fig:2}). 
In this case, the final state of the process contains two neutrinos 
(if they are Majorana particles) or a $\nu\bar\nu$ pair (if they are of Dirac 
nature), so in the Majorana case the antisymmetrization has to be carried 
out. However, just as in the case 
of the neutrino pair production in $e^+e^-\to Z^*\to \nu\nu$ reaction, 
no differences between the cross sections in the Dirac and Majorana cases 
arise in the limit $m_\nu\to 0$ if the final-state neutrinos are not 
detected. Thus, we are back to the situation one had 
for the neutrino pair production in $e^+e^-\to Z^*\to \nu\nu$ reaction:   
Dirac {\em vs.} Majorana neutrino nature 
could only be probed 
by detection of final-state neutrinos. 
CC detection shows that for negligibly small neutrino mass the 
Dirac/Majorana differences disappear; NC detection leads to the same conclusion 
if the final-state neutrinos in NC processes are not observed. 

Let us now study the NC detection process of fig.~\ref{fig:2} in 
more detail. As in the case of CC detection, it is instructive to examine the 
situation when the neutrino production and detection processes are separated by 
macroscopically large distances and therefore can be considered as independent.  
Let us first discuss neutrino detection in the Dirac case. 
The matrix elements describing the scattering of muon neutrinos and 
antineutrinos on electrons 
in one of the detectors are given, up to a constant factor, by  
\begin{align}
{\cal M}^\mathrm{D}_{\nu_\mu e}=\,
&{\cal J}_\mu \big[\bar{u}_{s_1'}(p_1')
\gamma^\mu(1-\gamma_5)u_{s_1}(p_1)\big]\,,
\label{eq:nueNC} \\
{\cal M}^\mathrm{D}_{\bar{\nu}_\mu e}=\,&
{\cal J}_\mu \big[\bar{v}_{s_1}(p_1)\gamma^\mu(1-\gamma_5)
v_{s_1'}(p_1')\big]
=\,
{\cal J}_\mu \big[\bar{u}_{s_1'}(p_1')
\gamma^\mu(1+\gamma_5)u_{s_1}(p_1)\big]\,. 
\label{eq:antinueNC}
\end{align}
Here $p_1,s_1$ and $p_1',s_1'$ 
denote the four-momenta and spins of the initial-state and 
final-state neutrinos or antineutrinos, and ${\cal J}_\mu$ is 
the convolution of the 
electron NC matrix element $j^\nu_e$ and the $Z^0$-boson propagator 
$D^Z_{\nu\mu}$: ${\cal J}_\mu=j^\nu_e D^Z_{\nu\mu}$. 
It is not possible to find out on a case-by-case basis 
whether a given neutrino detector has observed $\nu_\mu$ or  $\bar{\nu}_\mu$; 
however, if one of them detects $\nu_\mu$, the other will detect 
$\bar{\nu}_\mu$, and vice versa. Therefore, the number of scattering events 
in both detectors in a simultaneous detection experiment is proportional to
$\Big(\frac{\mathrm{d}\sigma_{\nu_\mu e}^\mathrm{D}}{\mathrm{d}T}+
\frac{\mathrm{d}\sigma_{\bar{\nu}_\mu e}^\mathrm{D}}{\mathrm{d}T}\Big)$, 
where $T$ is the kinetic energy of the recoil electron. 

Let us now consider the case of Majorana neutrinos. The matrix element of 
$\nu_\mu e$ scattering in one of the detectors is, up to a constant factor, 
\begin{align}
{\cal M}^{\rm M}_{\nu_\mu e}=
&{\cal J}_\mu
\big[\bar{u}_{s_1'}(p_1')\gamma^\mu(1-\gamma_5)
u_{s_1}(p_1)-
\bar{v}_{s_1}(p_1)\gamma^\mu(1-\gamma_5)
v_{s_1'}(p_1')\big]\nonumber \\
=&{\cal J}_\mu
\big[\bar{u}_{s_1'}(p_1')\gamma^\mu(1-\gamma_5)
u_{s_1}(p_1)-
\bar{u}_{s_1'}(p_1')\gamma^\mu(1+\gamma_5)
u_{s_1}(p_1)\big]\,,
\label{eq:nueMaj1}
\end{align}
and similarly for the other detector.%
\footnote{The neutrino NC matrix element here is 
actually that given in eq.~(\ref{eq:MajNeutral1}), where it has now been 
taken into account that $g_V=g_A$.}
The two terms in the square brackets in the second line sum up to the 
pure axial vector 
$-2\bar{u}_{s_1'}(p_1')\gamma^\mu\gamma_5u_{s_1}(p_1)$; 
however, for our purposes it will be more convenient to use 
the expression for ${\cal M}^{\rm M}_{\nu_\mu e}$ as it is given in the 
second line of eq.~(\ref{eq:nueMaj1}). 
Comparing~(\ref{eq:nueMaj1}) with eqs.~(\ref{eq:nueNC}) and 
(\ref{eq:antinueNC}), we find 
\be
{\cal M}^{\rm M}_{\nu_\mu e}={\cal M}^\mathrm{D}_{\nu_\mu e}-
{\cal M}^\mathrm{D}_{\bar{\nu}_\mu e}\,.
\label{eq:nueMaj2}
\ee
The squared matrix element of the scattering process can be written as 
$L_{\mu\nu}N^{\mu\nu}$, where 
$L_{\mu\nu}={\cal J}_\mu{\cal J}^*_\nu$, 
and the bilinear neutrino current products $N^{\mu\nu}$ depend on the 
neutrino nature and 
on the scattering process. In the Dirac case, one has, 
for $\nu_\mu e$ and $\bar{\nu}_\mu e$ scattering,   
\be
N_{\nu_\mu e}^{(\mathrm{D})\mu\nu}~=~
\big[\bar{u}_{s_1'}(p_1')\gamma^\mu(1-\gamma_5)u_{s_1}(p_1)\big]
\big[\bar{u}_{s_1}(p_1)\gamma^\nu(1-\gamma_5)u_{s_1'}(p_1')\big]\,,
\vspace*{-3mm}
\ee
\begin{align}
N_{\bar{\nu}_\mu e}^{(\mathrm{D})\mu\nu}~=~&
\big[\bar{v}_{s_1}(p_1)\gamma^\mu(1-\gamma_5)v_{s_1'}(p_1')\big]
\big[\bar{v}_{s_1'}(p_1')\gamma^\nu(1-\gamma_5)v_{s_1}(p_1)\big]\nonumber \\
~=~&\big[\bar{u}_{s_1'}(p_1')\gamma^\mu(1+\gamma_5)u_{s_1}(p_1)\big]
\big[\bar{u}_{s_1}(p_1)\gamma^\nu(1+\gamma_5)u_{s_1'}(p_1')\big]\,. 
\end{align}
In the Majorana case, 
we find 
\be
N_{\nu_\mu e}^{(\mathrm{M})\mu\nu}=N_{\nu_\mu e}^{(\mathrm{D})\mu\nu}+
N_{\bar{\nu}_\mu e}^{(\mathrm{D})\mu\nu}
-T^{\mu\nu}\,, 
\label{eq:nueMaj3}
\ee
where use has been made of the relation in eq.~(\ref{eq:nueMaj2}). The 
quantity $T^{\mu\nu}$ comes from the interference of 
the two terms in eq.~(\ref{eq:nueMaj2}) and is given by 
\begin{align}
T^{\mu\nu}~=~&
\big[\bar{u}_{s_1'}(p_1')\gamma^\mu(1-\gamma_5)u_{s_1}(p_1)\big]
\big[\bar{u}_{s_1}(p_1)\gamma^\nu
(1+\gamma_5)u_{s_1'}(p_1')\big]\nonumber& \\
& +\big[\bar{u}_{s_1'}(p_1')\gamma^\mu(1+\gamma_5)u_{s_1}(p_1)\big]
\big[\bar{u}_{s_1}(p_1)\gamma^\nu(1-\gamma_5)u_{s_1'}(p_1')\big] \nonumber& \\
~=~&-\mathrm{tr}\big\{\slashed{p}_1'\slashed{s}_1'\gamma^\mu\slashed{p}_1
\slashed{s}_1\gamma^\nu-m_\nu\big(
\slashed{p}_1'\slashed{s}_1'\gamma^\mu\gamma^\nu-\gamma^\mu\slashed{p}_1
\slashed{s}_1\gamma^\nu\big)\gamma_5
-m_\nu^2\gamma^\mu\gamma^\nu\big\}\,.&
\label{eq:T}
\end{align}
Here $s_1$ and $s_1'$ are the spin four-vectors of the incident and scattered 
neutrinos,  
\be
s_1^\mu=\left(
\frac{\vec{p}_1\cdot \vec{n}_1}{m_\nu},\,\vec{n}_1+\frac{(\vec{p}_1\cdot 
\vec{n}_1) \vec{p}_1}{m_\nu(E_1+m_\nu)} \right),
\label{eq:4spin}
\ee
with $\vec{n}_1$ being the unit spin vector of the incoming neutrino in its 
rest frame, and similarly for $s_1'^{\mu}$.
Next, we note that the detection of the neutrinos produced in 
$e^+e^-\to Z^*\to \nu\nu$ reaction 
is achieved through the measurement of electron recoil in $\nu_\mu e$ 
scattering, i.e.\ the scattered neutrinos in the final state are not observed.
Therefore, the summation over the spin $s_1'$ 
(which amounts to integration over $\vec{n}_1'$) has to 
be carried out and we find 
\be
\sum_{s'}T^{\mu\nu}=
\mathrm{tr}\big\{-m_\nu\gamma^\mu\slashed{p}_1
\slashed{s}_1\gamma^\nu\gamma_5
+m_\nu^2\gamma^\mu\gamma^\nu\big\}\,. 
\label{eq:Tsum}
\ee
Thus, the interference term $T^{\mu\nu}$ is suppressed for relativistic 
neutrinos at least as $m_\nu/E$. From 
eq.~(\ref{eq:nueMaj3}) we then find that for negligibly small neutrino mass 
the number of events in one neutrino detector in the Majorana case is 
proportional to the sum  
of the Dirac neutrino cross sections of $\nu_\mu e$ and $\bar{\nu}_\mu e$ 
scattering events. This has to be multiplied by a factor of two 
for two detectors, but on the other hand there is a factor 1/2 because of  
the identical nature of the two neutrinos in 
the reaction $e^+e^-\to Z^*\to \nu\nu$.  
The two factors compensate each other, and the total number of events in the 
simultaneous neutrino detection experiment remains 
proportional to $\Big(\frac{\mathrm{d}\sigma_{\nu_\mu e}^\mathrm{D}}{\mathrm{d}T}+
\frac{\mathrm{d}\sigma_{\bar{\nu}_\mu e}^\mathrm{D}}{\mathrm{d}T}\Big)$, 
with the same proportionality factor as in the Dirac neutrino case. 
Thus, also in the case of neutrino detection through NC processes, the 
difference between the cross sections for Dirac and Majorana neutrinos 
smoothly disappears in the limit $m_\nu\to 0$, with no need to drop any  
terms in the squared matrix element~by~hand. 

A key conclusion from the above calculation is that in the limit $m_\nu\to 0$ 
Majorana neutrinos produced or destroyed by their NC can 
be considered to be in states of definite chirality, despite the purely 
axial-vector form of the current. This is a consequence of the fact 
that the interference between the opposite chirality amplitudes vanishes 
in this limit. 

It is obvious that the confusion theorem holds also true when one of the 
neutrinos from $Z^*\to\nu\nu$ decay is detected through a CC process and 
another through a NC process. The CC interaction selects a neutrino of one 
predominant helicity, either positive or negative, depending on the process; 
such detection  modes satisfy the Practical Dirac-Majorana Confusion 
Theorem, as discussed in section~\ref{sec:1nu}. 
The other neutrino will then have the opposite predominant helicity, and its 
detection through a NC interaction will have, in the $m_\nu\to 0$ limit, 
the same cross section for Dirac and Majorana neutrinos, as discussed 
in section~\ref{sec:neutral}. 

Thus, irrespectively of the detection processes, in the case of negligibly 
small neutrino mass observation of both 
neutrinos produced in the reaction $e^+e^-\to Z^*\to 
\nu\bar{\nu}(\nu\nu)$ would result in the same cross section  
for Dirac and Majorana neutrinos. If at least one of the 
produced neutrinos escapes unobserved, the inherent summation over its spin 
leads to the same result. Our conclusion is obviously also valid for 
reactions of neutrino pair creation and annihilation, which are related to 
neutrino scattering processes by crossing symmetry.

\section{General proof} 
\label{sec:generalProof}

In refs.~\cite{Kim:2021dyj} and~\cite{Kim:2023iwz} it has been asserted that 
no general proof of the Practical Dirac-Majorana Confusion Theorem exists, 
and its validity had only been demonstrated for a few processes. This was  
a motivation for the authors to search for possible exceptions to this theorem. 
In our opinion, the very 
general argument~\cite{Case:1957zza,Li:1981um,Kayser:1981nw,Kayser:1982br} 
that the theorem should be universally valid within the Standard Model because 
both Dirac and Majorana neutrinos become Weyl neutrinos in the $m_\nu\to 0$ 
limit and the right-handed components of Dirac neutrinos are sterile, 
is by itself a sufficient proof of the confusion theorem. 
Still, some people prefer explicit proofs to general arguments; 
we therefore present here what we believe constitutes such a proof. 
    
Consider a generic neutrino process due to $n$ CC and $k$ NC interactions 
($n+k\ge 1$).%
\footnote{We do not include in our discussion neutrino oscillations, which are 
well known to be independent of Dirac {\it vs.} Majorana neutrino nature. Not 
included here are also processes caused by neutrino electromagnetic 
interactions; we comment on them at the end of this section.} 

\vspace*{2mm}
\noindent
I. {\em CC processes.}
As discussed in section~\ref{sec:1nu}, the chiral structure 
of CC interactions implies that in the limit $m_\nu\to 0$ the corresponding 
matrix elements for Dirac and Majorana neutrinos coincide, except possibly 
for antisymmetrization due to identical nature of some final-state neutrinos, 
which may in principle be different in the Dirac and Majorana cases and 
which will be discussed later on.

\vspace*{2mm}
\noindent
II. {\em NC processes.} These may be either of scattering or of neutrino 
pair production or annihilation type.  
In scattering processes one initial-state neutrino or antineutrino is 
destroyed and one is produced in the final state; in pair production and 
annihilation processes, a 
$\nu\bar{\nu}$ or $\nu\nu$ pair is either produced 
or annihilated by each NC interaction. 

\vspace*{1mm}
\noindent
(i) {\em   
NC scattering processes}. In the Dirac case, neutrino NC interaction has 
chiral structure, while for Majorana neutrinos, it is purely axial-vector. 
The analysis 
of Dirac/Majorana differences depends on the production mechanism of 
incident neutrinos in the Majorana case. There are two possibilities: 
\begin{itemize}
\item[(a)] An incident Majorana neutrino has been previously produced in a 
CC process and therefore participates in the process under 
consideration as a state of (nearly) definite chirality. 
Because axial-vector interactions do not flip chirality, the 
scattered neutrino will then be in the same chirality state. As shown in 
\cite{Kayser:1981nw,Kayser:1997hj,Hannestad:1997mi,Hansen:1997sk,Zralek:1997sa,
Czakon:1999ed} (see also section~\ref{sec:neutral} above), 
in this case the difference between the NC matrix elements in the Dirac and 
Majorana cases vanishes in the limit $m_\nu\to 0$. 

\item[(b)] 
An incident Majorana neutrino in the NC scattering process 
was produced in a NC pair production reaction. 
This case was considered in section~\ref{sec:NCdetection}. 
Being caused by a purely axial-vector interaction, 
this process in general leads to the production of a pair of neutrinos 
of no definite chirality. Indeed,  for Majorana neutrinos the NC matrix element 
is always given by a difference of two amplitudes of opposite chirality, 
regardless of whether neutrino scattering or pair-production is considered 
(cf.\ eqs.~(\ref{eq:MajNeutral1}) and (\ref{eq:MajNeutral2})). This is 
related to the fact that the NC contains one $\nu(x)$ and one $\bar{\nu}(x)$ 
field operator, and in the Majorana case each of these operators can both 
create and annihilate a neutrino. As was shown in  
section~\ref{sec:NCdetection}, these two terms do not interfere in the limit 
$m_\nu\to 0$ if the neutrino spins are not measured, which means that in this 
limit Majorana neutrinos originating from pair-production NC processes 
will also participate in the NC scattering process as 
states of definite chirality. As discussed in point (a) above, this means 
that to leading order in $m_\nu/E$ there will be no differences between the 
corresponding cross sections in the Dirac and Majorana neutrino cases. 

\end{itemize}

\vspace*{1mm}
\noindent
(ii) {\em NC neutrino pair production or annihilation.} The conclusion 
in section~\ref{sec:NCdetection} regarding the chiral 
nature of Majorana neutrinos born in NC pair-production processes applies 
not only to production of incident neutrinos, but also 
to pair-production of final-state neutrinos and to  
annihilation of initial-state ones in the generic process we are 
discussing here. As was stressed above, the differences 
between the Dirac and Majorana matrix elements disappear in the limit 
$m_\nu\to 0$ when neutrinos are in chiral states. 

\vspace*{2mm}
This exhausts all possible neutrino processes induced by CC and NC 
weak interactions. 
The only issue that has so far been left out of our general proof is the 
antisymmetrization of the amplitudes in the Dirac and Majorana cases. 
We consider it next. 

III. {\em Antisymmetrization}. For Dirac neutrinos, antisymmetrization should 
be done between each pair of 
neutrinos in the final state, and similarly for each pair of antineutrinos.%
\footnote{Recall that in this paper we consider all neutrinos to be 
of same flavor.}  
In the Majorana case, the antisymmetrization 
should in principle be carried out for each pair of produced neutrinos. 
As follows from the discussion above in this section, 
all finite-state neutrinos and/or antineutrinos in the  
generic process we discuss can be considered as emitted in states of 
definite chirality, irrespectively of their Dirac/Majorana nature and of 
whether they are produced by CC or NC interactions. 
We have demonstrated in section~\ref{sec:NCdetection} that 
to leading order in $m_\nu$ the interference between terms of opposite 
chiralities in the transition amplitude disappears, and that for this reason 
the antisymmetrization in the Majorana case 
to this order should only be done between pairs of neutrinos of 
same chirality. Recall now that Dirac neutrinos and antineutrinos are produced 
as states of, respectively, negative and positive chiralities.
Thus, to leading order in $m_\nu/E$, antisymmetrization should only be done 
in the Majorana case between the same neutrinos 
for which it should be performed  
in the Dirac case; all effects of ``additional'' antisymmetrization 
related to the Majorana nature of neutrinos will be suppressed at least as 
$m_\nu/E$. This completes our proof. 

Note that the cases when some of neutrinos are 
Dirac particles while others are of Majorana nature are obviously also 
covered by the generic case discussed above. 

We did not discuss here processes induced by electromagnetic interactions 
of neutrinos. The validity of the Practical Dirac-Majorana Confusion Theorem 
for such processes was proven in refs.~\cite{Li:1981um,Schechter:1981hw,
Kayser:1982br,Barr:1987ht,Musolf:1990sa} (see also~\cite{Giunti:2014ixa}). 

\section{Summary and conclusion} 
\label{sec:disc}

We have studied in detail the question of whether antisymmetrization of the 
amplitudes of the processes with more than one neutrino in the final state 
required by quantum statistics can help distinguish Dirac from Majorana 
neutrinos. We examined the existing claims in the literature 
that the differences between the cross sections of Dirac and Majorana 
neutrinos due to this antisymmetrization survive even for arbitrarily small 
but not exactly vanishing neutrino mass and demonstrated that they do not 
hold true. We also analyzed the implications of quantum statistics for generic 
neutrino processes and presented a general proof of the Practical 
Dirac-Majorana Confusion Theorem with the antisymmetrization issue 
explicitly taken into account. The key point in our analysis was the 
observation that for Majorana neutrinos chirality, which is nearly 
conserved for relativistic fermions, plays essentially the same role as 
lepton number conservation plays for Dirac neutrinos. 
Because the difference between chirality and 
(exactly conserved) helicity is suppressed 
as ${\cal O}(m_\nu/E)$, so are the differences between 
the observables in the Dirac and Majorana neutrino cases. 

Leaving aside lepton number violating processes whose probabilities are 
explicitly proportional to Majorana neutrino mass and therefore vanish when 
it goes to zero, there are two main differences between the matrix elements 
for Dirac and Majorana neutrinos: 

\begin{itemize}
\item[A.]
In the Dirac neutrino case, NC is chiral, whereas for Majorana neutrinos 
it is purely axial-vector;

\item[B.]
In the Dirac case, antisymmetrization of the amplitudes has to be done 
with respect to the interchange of each  
neutrino pair in the final state, and similarly for each antineutrino pair; 
at the same time, since for Majorana neutrinos there is no difference 
between neutrinos and antineutrinos, the antisymmetrization should be 
carried out with respect to the interchange of each pair of 
neutrinos in the final state.
\end{itemize}

It was proven long ago that for NC-induced neutrino scattering processes, 
such as $\nu_\mu e$ scattering 
or deuteron disintegration 
$\nu+d\to p+n+\nu$, the axial vector nature of the NC of Majorana neutrinos 
does not violate the confusion theorem provided that the incident 
neutrinos have been produced in a CC process and therefore are (nearly) 
chiral. This is because the axial-vector current, though does not 
project out states of definite chirality, does not flip the chirality 
either; the scattered neutrino is therefore essentially in the same chirality 
state as the incoming neutrino.  

The situation when the incident neutrino in a NC scattering process was 
previously produced in another NC process was for the first time considered in 
this paper. The subtlety is that in the Majorana case, due to the axial-vector 
nature of the NC, the produced neutrinos are in general born in states of no 
definite chirality. Rather, the production amplitude is a coherent sum 
of terms corresponding to emission of neutrinos of opposite chirality. We have 
demonstrated that the interference of these two terms is suppressed 
for relativistic neutrinos at least as $m_\nu/E$, and therefore to leading 
order it can be neglected. This means that in the limit $m_\nu/E\to 0$ Majorana 
neutrinos produced (destroyed) in pair-production (annihilation) processes 
can also be considered to be in states of definite chirality. Thus, the 
peculiarity of Majorana neutrinos mentioned in point A above does not 
lead to any exception to the practical confusion theorem. \\
\hspace*{6mm}
The conclusion that to leading order in $m_\nu/E$ neutrinos can always 
be considered as being in states of definite chirality regardless of their 
origin, Dirac/Majorana nature and of the process in which they participate 
has also direct bearing on the antisymmetrization issue mentioned in point B. 
Because in the limit $m_\nu\to 0$ 
chirality is a good quantum number, it plays in this limit essentially the 
role of lepton number for Majorana neutrinos. 
The antisymmetrization in the Majorana case should then only be done 
for those neutrinos for which it should be done in the 
Dirac neutrino case. All corrections to this rule are suppressed at least as 
$m_\nu/E$ and therefore do not violate the practical confusion theorem. \\
\hspace*{6mm}
In conclusion, we have shown that the Practical Dirac-Majorana Confusion 
Theorem applies without restrictions within the Standard Model, and 
in particular that effects of quantum statistics do not lead to any 
exceptions to it.\\ 
\hspace*{6mm}The results of our study are in agreement with 
\vspace*{1.5mm}
Hinchliffe's 
rule~\cite{Peon:1988kx}.\\
%
\hspace*{6mm}
{\em Notes added.} After this paper was submitted to the e-print archive, our 
attention was drawn to ref.~\cite{Marquez:2023rpc}, where the leptonic 
radiative four-body decay $\ell\to\ell'\nu_\ell\bar{\nu}_{\ell'}\gamma$ was 
considered. Although this process is different from that studied 
in~\cite{Kim:2021dyj}, their back-to-back kinematics is essentially the same.   
The authors of~\cite{Marquez:2023rpc} 
came to the conclusion that the integration over the unobservable 
angle $\theta$ between the neutrino and $\ell'\gamma$ directions eliminates 
all the Dirac/Majorana differences in the differential decay rates 
to leading order in $m_\nu/E$, and that the same applies to the $B^0$-decay 
process considered in~\cite{Kim:2021dyj}, where $\theta$ is the angle 
between the neutrino and muon directions. This result is in 
full agreement with that in our section~\ref{sec:B0decay}. It should be 
stressed, however, that in the case when no integration over $\theta$ is 
carried out, the differences between neutrino energy and angular 
distributions of Dirac and Majorana neutrinos in the limit $m_\nu/E\to 0$,  
shown in fig.~5 of ref.~\cite{Marquez:2023rpc}, are actually unobservable 
within the Standard Model.\ As was demonstrated in the general case in 
section~\ref{sec:generalProof} of our paper and also discussed in detail for 
the special case of NC processes in sections~\ref{sec:CCdetection} and 
\ref{sec:NCdetection}, to analyze the 
dependence of these observables on neutrino nature one has to include the 
neutrino detection  processes in the consideration, and this would eliminate 
the Dirac/Majorana differences in the limit $m_\nu/E\to 0$. \\
\hspace*{6mm}
In a recent note~\cite{Kim:2024xof} the authors of~\cite{Kim:2021dyj} 
attempted at a rebuttal of our criticism of their paper. 
However, they did not answer our critical remarks on the limiting procedure 
that was used in~\cite{Kim:2021dyj} to derive the expression for the angle 
$\theta$ (which actually led to their main conclusions) and 
just reiterated that their analysis was thorough and accurate. We 
maintain the validity of our criticism of the results of 
ref.~\cite{Kim:2021dyj}. 
\enlargethispage{0.5cm}

\acknowledgments
We are grateful to Alexei Smirnov for numerous useful discussions.\ EA is 
grateful to C.S.~Kim and Dibyakrupa Sahoo for discussions concerning 
refs.~\cite{Kim:2021dyj} and~\cite{Kim:2023iwz}. We thank J.~M.~M\'{a}rquez 
for drawing our attention to ref.~\cite{Marquez:2023rpc}. 
All diagrams in this work were generated with
\texttt{JaxoDraw}~\cite{Binosi:2003yf}.

\bibliographystyle{JHEP}
\bibliography{bibliography}

\end{document}